\documentclass{article}

\usepackage[preprint]{neurips_2025}

\usepackage[utf8]{inputenc} 
\usepackage[T1]{fontenc}    
\usepackage{hyperref}       
\usepackage{url}            
\usepackage{booktabs}       
\usepackage{amsfonts}       
\usepackage{nicefrac}       
\usepackage{microtype}      
\usepackage{xcolor}         
\usepackage[utf8]{inputenc}
\usepackage{amsmath}
\usepackage{amsfonts}
\usepackage{graphicx}
\usepackage{multirow}

\title{HORM: A Large Scale Molecular \underline{H}essian Database for \underline{O}ptimizing \underline{R}eactive \underline{M}achine Learning Interatomic Potentials}

\author{%
  Taoyong Cui\thanks{Equal contribution.} \\
  Deep Principle Inc.\\
  \texttt{cuitaoyong@deepprinciple.com} \\
  \And
  Yunhong Han\footnotemark[1] \\
  Deep Principle Inc. \\
  \texttt{yunhonghan@deepprinciple.com} \\
  \And
  Haojun Jia\thanks{Corresponding Author.}\\
  Deep Principle Inc.  \\
  \texttt{\texttt{haojunjia@deepprinciple.com}} \\
  \And
  Chenru Duan\footnotemark[2] \\
  Deep Principle Inc. \\
  \texttt{duanchenru@gmail.com} \\
  \And
  Qiyuan Zhao\footnotemark[2] \\
  Deep Principle Inc. \\
  \texttt{zhaoqiyuan@deepprinciple.com} \\
}

\begin{document}

\maketitle

\begin{abstract}
Transition state (TS) characterization is central to computational reaction modeling, yet conventional approaches depend on expensive density functional theory (DFT) calculations, limiting their scalability. Machine learning interatomic potentials (MLIPs) have emerged as a promising approach to accelerate TS searches by approximating quantum-level accuracy at a fraction of the cost. However, most MLIPs are primarily designed for energy and force prediction, thus their capacity to accurately estimate Hessians, which are crucial for TS optimization, remains constrained by limited training data and inadequate learning strategies. This work introduces the Hessian dataset for Optimizing Reactive MLIP (HORM), the largest quantum chemistry Hessian database dedicated to reactive systems, comprising 1.84 million Hessian matrices computed at the $\omega$B97x/6-31G(d) level of theory. To effectively leverage this dataset, we adopt a Hessian-informed training strategy that incorporates stochastic row sampling, which addresses the dramatically increased cost and complexity of incorporating second-order information into MLIPs. Various MLIP architectures and force prediction schemes trained on HORM demonstrate up to 63\% reduction in the Hessian mean absolute error and up to 200 times increase in TS search compared to models trained without Hessian information. These results highlight how HORM addresses critical data and methodological gaps, enabling the development of more accurate and robust reactive MLIPs for large-scale reaction network exploration.
\end{abstract}

\section{Introduction}
Computational transition state (TS) characterization is essential for elucidating reaction mechanisms, differentiating between competing reaction pathways, and predicting reaction kinetics and thermodynamics, making it an indispensable tool in computational chemistry for a wide range of applications.\cite{das2019insights,kang2019glucose,zhao2023deep,zhao2023thermally,wen2023chemical,steiner2024human} Traditional computational TS search relies on costly density functional theory (DFT) calculations to evaluate energy and forces across large reactive spaces, along with Hessian calculations for saddle point optimization.\cite{zhao2021YARP,zhao2022YARP2,young2021autode} However, the exponential growth of chemical space in modern drug discovery and materials science makes such costly calculations increasingly inadequate for meeting the demands of large-scale reaction predictions.\cite{duan2024react} To address the high computational demands, machine learning (ML) approaches have emerged as powerful accelerators by significantly reducing dependence on DFT calculations.\cite{wen2023chemical,duan2023dec,duan2024react,zhang2024may,vadaddi2024graph} One of the most promising tools is machine learning interatomic potentials (MLIPs), which accurately characterize potential energy surfaces (PES) at low computational cost.\cite{fedik2022sep,kaser2023neural} By enabling efficient evaluation of energy landscapes, interatomic forces, and Hessian matrices, MLIPs naturally integrate with physics-based transition state search algorithms.

MLIPs have emerged as powerful tools for predicting energies and forces with quantum-mechanical accuracy.\cite{Deringer2019,poltavsky2021jul,unke2021aug,Nandy2021Aug,Kocer2022,anstine2023machine,kaser2023neural} 
While state-of-the-art architectures such as EquiformerV2,\cite{liao2024equiformerv} MACE,\cite{maceoff,Batatia2022mace} and Orb \cite{neumann2024Oct} have achieved remarkable accuracy on equilibrium systems, extending their applicability to reactive systems involving bond-breaking and bond-forming remains a significant challenge due to the broader chemical diversity and more complex regions of the potential energy surface.
Schreiner et al. showed that PaiNN can identify transition states across diverse reactions when trained on the Transition1x dataset.\cite{NeuralNEB} 
Yuan et al. further demonstrated that fine-tuning an equivariant message-passing neural network on Transition1x can enhance TS optimization, even without access to second-order (Hessian) information.\cite{yuan2024analytical}
More recently, Anstine et al. introduced AIMNet2-Pd, expanding the chemical domain of reactive MLIPs from main-group elements to transition metal catalysis by training on a diverse dataset of Pd-catalyzed cross-coupling reactions.\cite{anstine2025transferable}
However, a recent benchmark study evaluating universal MLIPs in an end-to-end TS search workflow reveals that many models—especially models that directly predict forces—struggle with transition state optimization due to inaccurate and even asymmetric Hessian predictions, a limitation rooted in training solely on energies and forces without access to second-order information.\cite{zhao2025harnessing}

The common exclusion of Hessian matrices in MLIP development stems from two primary limitations: the scarcity of off-equilibrium Hessian data due to their high computational cost, and methodological challenges in incorporating second-order information efficiently during training. In this work, we introduce HORM (Hessian dataset for Optimizing Reactive MLIP), a large-scale quantum chemistry database specifically designed to address these issues and enhance reactive MLIP development, which comprises 1,836,206 Hessian matrices computed at the $\omega$B97x/6-31G(d) level of theory. By systematically incorporating Hessian information, HORM enables more effective training paradigms that improve second-derivative accuracy and transition state identification. Notably, models that directly predict forces, which are typically prone to asymmetric and physically inconsistent Hessian estimates, benefit significantly from this enhancement.

This paper makes the following contributions:

\begin{itemize}
    \item We present HORM, the largest quantum chemistry dataset tailored to reactive systems, comprising 1.84 million off-equilibrium Hessian matrices computed at the $\omega$B97x/6-31G(d) level of theory. HORM expands the availability of second-order information by near two orders of magnitude compared to existing datasets (e.g., Hessian-QM9: 41,645 equilibrium Hessians).

    \item We propose and systematically benchmark Hessian-informed training strategies that augment standard energy-force objectives with second-order loss terms. These strategies significantly enhance Hessian accuracy and enforce symmetry, particularly benefiting models that directly predict forces.

    \item We develop reactive MLIPs that achieve substantial gains in TS search performance. Models trained on HORM achieve up to 200 times higher TS search success rates, underscoring the essential role of second-order information in enabling robust, end-to-end TS prediction workflows.  
\end{itemize}

\section{Preliminaries and Related Works}
\subsection{Machine Learning Interatomic Potentials}
MLIPs are models that learn to predict atomic interactions from quantum mechanical data, enabling efficient approximations of potential energy surfaces. Given an atomic configuration—defined by the elements and positions of atoms—MLIPs predict the total energy of the system and the corresponding atomic forces. MLIP models typically fall into two categories based on their force prediction strategy:

\begin{itemize}
    \item \textbf{Autograd-based models}: These models predict the total energy and compute forces as the negative gradient with respect to atomic positions: $\mathbf{F} = -\nabla_{\mathbf{r}} E$. This formulation ensures energy conservation and enforces physically meaningful symmetry in the force predictions. However, backpropagation through the energy function can increase training and inference costs.
    
    \item \textbf{Direct-force models}: These models treat energy and force prediction as separate tasks, directly learning atomic force vectors without computing them as gradients of the predicted energy. While this approach significantly enhances computational efficiency, it produces non-conservative forces and can introduce physical inconsistencies.\cite{bigi2024darkforces} Such discrepancies may degrade performance in downstream applications—particularly molecular dynamics simulations and transition state optimizations—where energy conservation and force symmetry are essential for accuracy and stability.
    
\end{itemize}

In this work, we implement three MLIPs, namely AlphaNet \cite{yin2025alphanet}, LEFTNet,\cite{leftnet} and EquiformerV2 \cite{liao2024equiformerv}, for TS search. LEFTNet is a SE(3)-equivariant graph neural network that incorporates efficient encoding of local 3D substructures and frame transitions. It has two versions: a direct-force model (LeftNet-df) and an autograd-based variant (LeftNet), the latter of which achieves the best performance in TS search tasks.\cite{zhao2025harnessing} AlphaNet is a local-frame-based SE(3)-equivariant autograd-based model, striking an excellent balance between computational efficiency and accuracy compared to existing models. EquiformerV2 is a Transformer-based architecture that leverages the equivariant spherical channel network (eSCN)\cite{passaro2023reducingso3convolutionsso2} convolutions to handle higher-degree tensors, enabling efficient E(3)-equivariance. As a representative direct-force model, EquiformerV2 has demonstrated excellent accuracy in both energy and force prediction.

\subsection{Hessian Matrix}
The Hessian matrix \(\mathbf{H} \in \mathbb{R}^{3N \times 3N}\) is a key quantity in molecular modeling, representing the second derivative of the potential energy \(U\) with respect to atomic coordinates \(\mathbf{r}\). Mathematically, it is expressed as:

\begin{equation}
    \mathbf{H} = \nabla_{\mathbf{r}}^2 U = -\frac{\partial \mathbf{F}}{\partial \mathbf{r}},
\label{eq:hessian}
\end{equation}

where \(\mathbf{F}\) denotes the atomic forces. The Hessian matrix provides critical information about the curvature of the potential energy surface, enabling the characterization of molecular stability and vibrational modes. In the context of TS search, the Hessian plays a pivotal role: first-order saddle point localization inherently depends on the local curvature of the energy landscape. Accurate Hessian information is typically required to guide the search direction and distinguish between minima and saddle points, ultimately improving the reliability and efficiency of TS optimization.

\subsection{Transition State Search}
TS search algorithms aim to locate first-order saddle points on potential energy surfaces, which represent the highest energy configurations along the minimal energy pathways between reactants and products. An accurate and automated TS search method typically involves two main steps: initial guess generation and refinement.\cite{zhao2021YARP} The first step generates an initial guess solely from reactant and product information. String-based methods, such as the Nudged Elastic Band (NEB)\cite{CINEB} and the Growing String Method (GSM),\cite{zimmerman2013GSM} are powerful tools for this purpose. Once a TS initial guess is obtained, it can be further refined using local Hessian-based optimization, while maintaining connection to the target reaction path. Convergence is determined by the presence of a single negative Hessian eigenvalue and small gradient norms. Following TS identification, Intrinsic Reaction Coordinate (IRC) calculations trace the complete reaction pathway through mass-weighted steepest descent in both forward (TS to products) and backward (TS to reactants) directions. If the two endpoints match the input reactants and products, the TS is called intended, indicating a successful TS search.
 
\subsection{Reactive Datasets}
\textbf{Transition1x.} The Transition1x dataset is one of the most widely used datasets for training reactive MLIPs.\cite{schreiner2022transition1x} It comprises 9.6 million molecular configurations sampled along and near reaction pathways, with energies and forces computed at the $\omega$B97x/6-31G(d) level of theory. The dataset was generated using the NEB method applied to 10,013 diverse organic reactions, with intermediate structures retained throughout the optimization process. It provides a comprehensive resource for studying reaction mechanisms by offering a rich variety of molecular geometries.

\textbf{RGD1.} The Reaction Graph Depth 1 (RGD1) dataset contains 950k organic reactions (177k computed at DFT level) involving C, H, O, and N atoms, with molecules containing up to 10 heavy (non-hydrogen) atoms.\cite{zhao2023comprehensive} Each reaction features at least one validated TS along with activation energy, heat of reaction, reactant and product geometries, and vibrational frequencies. The dataset is provided at the GFN2-xTB and B3LYP-D3/TZVP levels of theory, with a subset recalculated at higher levels (CCSD(T)-F12/cc-pVDZ-F12) to assess relative errors.

\subsection{Hessian Datasets}
Most current MLIP datasets primarily focus on energy and force information, with Hessian matrix data being relatively rare. One of the few publicly available Hessian datasets is the \textbf{Hessian-QM9} database, which contains equilibrium configurations and DFT-computed Hessian matrices for 41,645 molecules derived from the QM9 dataset \cite{williams2025hessian}. The calculations were performed at the $\omega$B97x/6-31G(d) level of theory, and the dataset represents a limited set of molecules in their equilibrium states. While valuable for improving the prediction of molecular vibrational properties, this database's reliance on equilibrium configurations limits its applicability to reactive systems. This gap underscores the need for more diverse Hessian datasets that include a broader range of molecular geometries, especially those relevant to reaction mechanisms and transition state optimization.


\subsection{Hessian Prediction}
\textbf{Equilibrium Structures.} Recent studies have explored various methodologies for predicting Hessian matrices of equilibrium structures. Hessian-QM9 leverages an E(3)-equivariant message-passing neural network trained on Hessians to enhance the prediction of molecular vibrational properties.\cite{rodriguez2025does} DetaNet introduces a partitioned tensor framework that decomposes the Hessian into atomic ($H_{ii}$) and interatomic ($H_{ij}$) components, enabling a modular and scalable prediction scheme.\cite{zou2023deep} Building on this, EnviroDetaNet incorporates environmental context into the architecture, resulting in a 42\% improvement in Hessian prediction accuracy and enabling more precise spectral predictions.\cite{xu2024envirodetanet}

\textbf{Non-equilibrium structures.}
For predicting Hessians of non-equilibrium molecular structures, existing approaches primarily rely on computing derivatives of predicted forces. A recent advancement by Amin et al.\cite{amin2025towards} introduces a knowledge distillation framework in which a MLIP student model learns Hessian representations from general-purpose foundation models (e.g., MACE-OFF23\cite{maceoff}) acting as teacher models. This approach enhances energy and force predictions for a variety of student models, despite the fact that the Hessians are not directly derived from DFT calculations. Meanwhile, TS optimization studies have underscored the crucial role of Hessians derived from MLIP models in ensuring reliable saddle point convergence.\cite{yuan2024analytical,zhao2025harnessing} These studies demonstrate that second-order information, which is implicitly captured in force predictions, can significantly assist in TS searches. However, a key limitation of these methods is that they do not explicitly incorporate Hessian supervision during training, thus restricting the accuracy and physical symmetry of the predicted Hessians.

In this work, we adopt a strategy inspired by Amin et al.\cite{amin2025towards}, where random columns of the Hessian matrix are learned to reduce computational overhead. This targeted supervision strikes an effective balance between computational cost and accuracy, leading to simultaneous improvements in energy and force predictions, Hessian fidelity, and TS search success rates.

\section{HORM Dataset}
\subsection{Dataset Composition}
The geometries in the HORM dataset are sampled from two reactive datasets, Transition1x and RGD1. To ensure consistency in the level of theory, all sampled geometries were recomputed by GPU4PYSCF\cite{wu2024gpu4pyscf} at the $\omega$ B97X/6-31G* level of theory to obtain energies, forces, and Hessian matrices.

\textbf{Transition1x.} 
Among the 10,073 reactions in Transition1x, we adopt a previously established data split based on reaction identities, assigning 9,000 reactions to the training set and the remaining 1,073 to the validation set.\cite{duan2024react} From these, 1,725,362 geometries corresponding to the training reactions and 50,844 geometries from the validation reactions are included in the final dataset, representing 20\% and 5\% of the available geometries from each split, respectively.

\textbf{RGD1.} To sample non-equilibrium geometries from the RGD1 dataset, we leverage reaction pathways generated via IRC calculations performed at the GFN2-xTB level of theory. From approximately 950,000 available reactions, we randomly selected 80,000 and sampled up to 15 geometries per reaction along their IRC trajectories. From this pool, 60,000 geometries were randomly chosen to constitute the final RGD1 subset.

\subsection{Dataset Analysis}
\begin{figure}[t]
    \centering
    \includegraphics[width=0.9\textwidth]{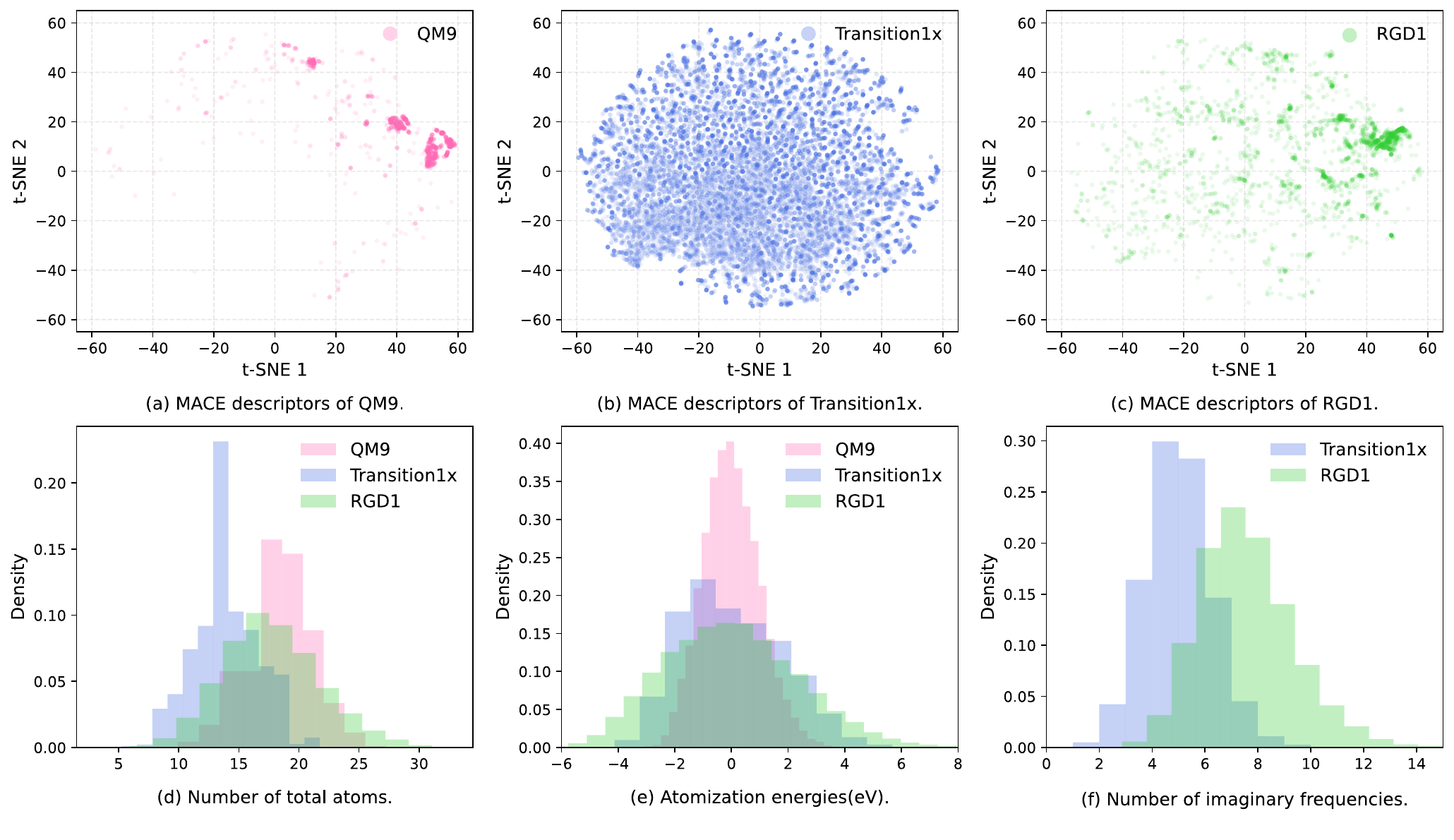}
    \caption{Molecular structure and property distributions of Hessian datasets. (a–c) t-SNE visualizations of MACE-OFF23-medium embeddings for 5\% of sampled geometries from Hessian-QM9, HORM-Transition1x, and HORM-RGD1. (d–f) Distributions of number of total atoms, atomization energies, and number of imaginary frequencies for Hessian-QM9, HORM-Transition1x, and HORM-RGD1.}
    \label{fig:dataset}
\end{figure}

\textbf{Chemical space coverage.} 5\% of geometries from Hessian-QM9, HORM-Transition1x, and HORM-RGD1 are sampled and embedded using the pre-trained MACE-OFF23 model\cite{maceoff} for t-SNE visualization (Fig. \ref{fig:dataset}a–c). Compared to Hessian-QM9, HORM spans a significantly broader region of chemical space. This expanded diversity is primarily attributed to the HORM-Transition1x subset, which contains approximately 40 times more molecular geometries than Hessian-QM9 and includes a wide variety of non-equilibrium structures. The HORM-RGD1 subset forms a distinct distribution with minimal overlap with HORM-Transition1x (Fig. \ref{fig:dataset}c). Its proximity to Hessian-QM9 in the embedded space—attributable to its larger molecular sizes—adds complementary diversity to HORM, thereby broadening its chemical coverage and improving its utility for benchmarking model generalization across a wide range of chemical environments.

\textbf{Properties distribution.} 
Compared to Hessian-QM9, the data points in the HORM dataset are, on average, smaller in molecular size—with Transition1x limited to molecules containing up to seven heavy atoms—but show a markedly broader distribution in the HORM-RGD1 subset (Fig. \ref{fig:dataset}d). The significantly wider range of atomization energies in HORM (Fig. \ref{fig:dataset}e) reflects its coverage of a much larger portion of the potential energy surface (PES), in contrast to the equilibrium-only structures of Hessian-QM9. The distribution of imaginary vibrational frequencies (Fig. \ref{fig:dataset}f) further highlights HORM's extensive sampling of non-equilibrium states. Notably, HORM-RGD1 exhibits even broader distributions than HORM-Transition1x in both atomization energies and the number of imaginary frequencies, indicating greater deviations from equilibrium and offering a more challenging benchmark for evaluating model generalization.

\section{Model Training}
An energy and force (E-F) loss function that combines errors in both energy and force predictions is commonly used for training MLIPs. The total loss function is defined as:

\begin{equation}
    \mathcal{L}_{\text{EF}} = \alpha \| E - \hat{E} \| + \frac{\beta}{N} \sum_{i=0}^{N-1} \left\| \mathbf{F}_{i} - \left( -\frac{\partial \hat{E}}{\partial \mathbf{r}_{i}} \right) \right\|,
    \label{eq:EF}
\end{equation}
where N is the total number of atoms and $\alpha$ and $\beta$ are the weights assigned to the energy and force losses, respectively.

Given a dataset $\mathcal{D} = \{(\mathbf{z}_i, \mathbf{r}_i, U_i, \mathbf{F}_i, \mathbf{H}_i)\}_{i=1}^N$ consisting of $N$ molecular structures with atomic numbers, positions, and DFT-calculated energy, force, and Hessian matrix, the MLIP ($\phi$) is trained to simultaneously match all three DFT-computed quantities. 

To mitigate the $\mathcal{O}(N^2)$ computational cost of full Hessian computation via automatic differentiation, we implement a stochastic row-sampling strategy. In each training epoch, $s$ rows $\mathcal{J}_i = \{j_1, \dots, j_s\} \subset \{1, \dots, 3N\}$ are randomly selected for each molecular structure in each epoch, where each index corresponds to an atomic coordinate. This sampling reduces the Hessian computation complexity to $\mathcal{O}(s)$ while maintaining learning efficacy. The modified loss function becomes:

\begin{equation}
    \mathcal{L}(\phi) = \mathcal{L}_{\text{EF}}(\phi) + \gamma\mathbb{E}_{\mathcal{J}_i \sim \mathcal{U}_s} \left( \frac{1}{s} \sum_{j \in \mathcal{J}_i} \left\| \mathbf{H}_i^{(j)} + \partial F_\phi^{(j)}(\mathbf{z}_i, \mathbf{r}_i)/ \partial \mathbf{r}  \right\|_2 \right) ,
\label{eq:EFH}
\end{equation}

where $\mathcal{U}_s$ denotes a uniform distribution (1,3N) over subsets of $s$ rows from the Hessian, $\gamma$ is the weight of the Hessian loss, $\mathbf{H}_i^{(j)}$ is the $j$-th row of the reference Hessian of molecule $i$, and $F_\phi^{(j)}$ is the MLIP-predicted force component.

Row extraction is efficiently performed via vector-Jacobian products (VJPs) \cite{amin2025towards}: given a MLIP force $\mathbf{F} : \mathbb{R}^{3N} \rightarrow \mathbb{R}^{3N}$ with Jacobian $\mathbf{H}$, selected rows $\mathbf{P}_{\mathcal{J}} \mathbf{H} \in \mathbb{R}^{s \times 3N}$ are computed using batched VJPs $\mathbf{v}^\top \partial_{\mathbf{r}} \mathbf{F}$, where $\mathbf{P}_{\mathcal{J}} = [\mathbf{e}_{j_1}, \dots, \mathbf{e}_{j_s}]^\top$ contains one-hot vectors $\mathbf{v}_k$. A \texttt{vmap}-optimized implementation further accelerates this process by avoiding explicit construction of the full Hessian matrix.


\section{Benchmarks and Evaluation}
\subsection{Energy, Force and Hessian Prediction}
This experiment evaluates different training strategies for energy and force prediction while assessing model generalization. We investigate the effect of incorporating Hessian supervision by benchmarking models trained with energy, force, and Hessian losses (E-F-H, Eq.\ref{eq:EFH}) against those trained with only energy and force losses (E-F, Eq.\ref{eq:EF}). All models are trained on the HORM-Transition1x training set and evaluated on both the in-distribution (ID) Transition1x validation set and the out-of-distribution (OOD) HORM-RGD1 subset.

In-distribution performance (Table \ref{tab:model-performance}) reflects the models' ability to learn from the training data. For both autograd and direct-force architectures, incorporating Hessian supervision consistently improves performance across nearly all evaluation metrics. In autograd-based models, the addition of Hessian loss reduces the energy MAE by up to 25\%, while resulting in minimal changes in the force MAE. However, it significantly improves second-order properties: Hessian and corresponding eigenvalue MAEs decrease by 59\% and 78\%, respectively. 

Direct-force models benefit more substantially from Hessian supervision, especially in the case of EquiformerV2, which achieves MAE reductions of 58\% in energy, 24\% in force, 97\% in Hessian, and 99\% in eigenvalue predictions. These results highlight the value of incorporating second-order information in directly enhancing the learned PES. Crucially, even though stochastic row sampling avoids full Hessian reconstruction, adding explicit Hessian loss markedly improves the symmetry of predicted Hessians—reducing asymmetry error (Eq. \ref{eq:asymmertry_value}) by up to 94\% (Table \ref{tab:asymmetry_comparison}). This leads to more physically consistent Hessian matrices, which are essential for robust and reliable TS optimization.

\begin{table}[t]
\centering
\renewcommand{\arraystretch}{1.6} 
\caption{\textbf{Comparison of errors for energy, force, and Hessian predictions on the HORM-Transition1x validation set.} Mean absolute errors are shown in the center of each column with median shown in parathensis. Bold values highlight the best-performing model in each task, separately for autograd-based and direct-force categories.}
\resizebox{\textwidth}{!}{ 
\begin{tabular}{llcccc}
\hline
\multirow{2}{*}{Type} & \multirow{2}{*}{Model} & Energy & Force & Hessian & Eigenvalues \\
 & & (eV) & (eV/\AA) & (eV/\AA$^2$) & (eV/\AA$^2$)\\
\hline
\multirow{4}{*}{Autograd-based} & AlphaNet (E-F) & 0.044(0.028) & 0.040(0.026) & 0.433(0.344) & 0.038(0.030) \\
 & AlphaNet (E-F-H)& \textbf{0.034(0.013)} & 0.040(0.026) & 0.303(0.215) & 0.024(0.016) \\
 & LEFTNet (E-F) & 0.047(0.033) & 0.037(0.025) & 0.366(0.302) & 0.032(0.026) \\
 & LEFTNet (E-F-H)& 0.035(0.017) & \textbf{0.036(0.025)} & \textbf{0.151(0.107)} & \textbf{0.007(0.004)} \\
\hline
\multirow{4}{*}{Direct-force} & LEFTNet-df (E-F) & 0.054(0.033) & 0.029(0.019) & 1.648(1.503) & 0.128(0.125) \\
 & LEFTNet-df (E-F-H)& 0.050(0.027) & 0.044(0.030) & 0.197(0.143) & 0.007(0.004) \\
 & EquiformerV2 (E-F) & 0.045(0.026) & 0.021(0.014) & 2.231(2.071) & 0.214(0.218) \\
 & EquiformerV2 (E-F-H)& \textbf{0.019(0.011)} & \textbf{0.016(0.010)} & \textbf{0.075(0.047)} & \textbf{0.003(0.001)} \\
\hline
\end{tabular}
}
\label{tab:model-performance}
\end{table}

\begin{table}[t]
\centering
\renewcommand{\arraystretch}{1.6} 
\caption{\textbf{Comparison of errors for energy, force, and Hessian predictions on the HORM-RGD1 subset.} Mean absolute errors are shown in the center of each column with median shown in parathensis. Bold values highlight the best-performing model in each task, separately for autograd-based and direct-force categories.}
\resizebox{\textwidth}{!}{ 
\begin{tabular}{llcccc}
\hline
\multirow{2}{*}{Type} & \multirow{2}{*}{Model} & Energy & Force & Hessian & Eigenvalues \\
 & & (eV) & (eV/\AA) & (eV/\AA$^2$) & (eV/\AA$^2$)\\
\hline
\multirow{4}{*}{Autograd-based} & AlphaNet (E-F) & 0.257(0.182) & 0.151(0.131) & 0.515(0.456) & 0.053(0.049) \\
 & AlphaNet (E-F-H) & 0.259(0.184) & 0.148(0.128) & 0.415(0.360) & 0.040(0.036) \\
 & LEFTNet (E-F) & 0.242(0.174) & 0.132(0.114) & 0.426(0.379) & 0.042(0.039) \\
 & LEFTNet (E-F-H) & \textbf{0.226(0.159)} & \textbf{0.130(0.112)} & \textbf{0.244(0.201)} & \textbf{0.015(0.012)} \\
\hline
\multirow{4}{*}{Direct-force} & LEFTNet-df (E-F) & 0.322(0.244) & 0.146(0.128) & 0.979(0.854) & 0.050(0.045) \\
 & LEFTNet-df (E-F-H) & 0.304(0.224) & 0.142(0.122) & 0.290(0.243) & 0.013(0.011) \\
 & EquiformerV2 (E-F) & 0.243(0.171) & 0.111(0.087) & 1.224(1.047) & 0.106(0.098) \\
 & EquiformerV2 (E-F-H) & \textbf{0.133(0.089)} & \textbf{0.056(0.038)} & \textbf{0.092(0.071)} & \textbf{0.003(0.003)} \\
\hline
\end{tabular}
}
\label{tab:rgd1-performance}
\end{table}

Out-of-distribution performance (Table \ref{tab:rgd1-performance}) evaluates the models’ generalization capabilities on unseen data. Similar patterns to those found in the in-distribution setting are observed. Autograd-based models show limited improvement in energy and force predictions but achieve notable gains in second-order properties, with Hessian and eigenvalue MAEs reduced by up to 43\% and 64\%, respectively. Among all models, the EquiformerV2 E-F-H variant—a representative direct-force architecture—not only achieves the largest reductions in prediction error (45\%, 50\%, 93\%, and 97\% for energy, force, Hessian, and eigenvalue, respectively), but also emerges as the best-performing model overall. Moreover, the consistently larger improvements in out-of-distribution performance compared to in-distribution performance underscore the critical role of incorporating Hessian supervision during training in enhancing model generalization.

\begin{figure}[htbp]
    \centering
    \includegraphics[width=0.8\textwidth]{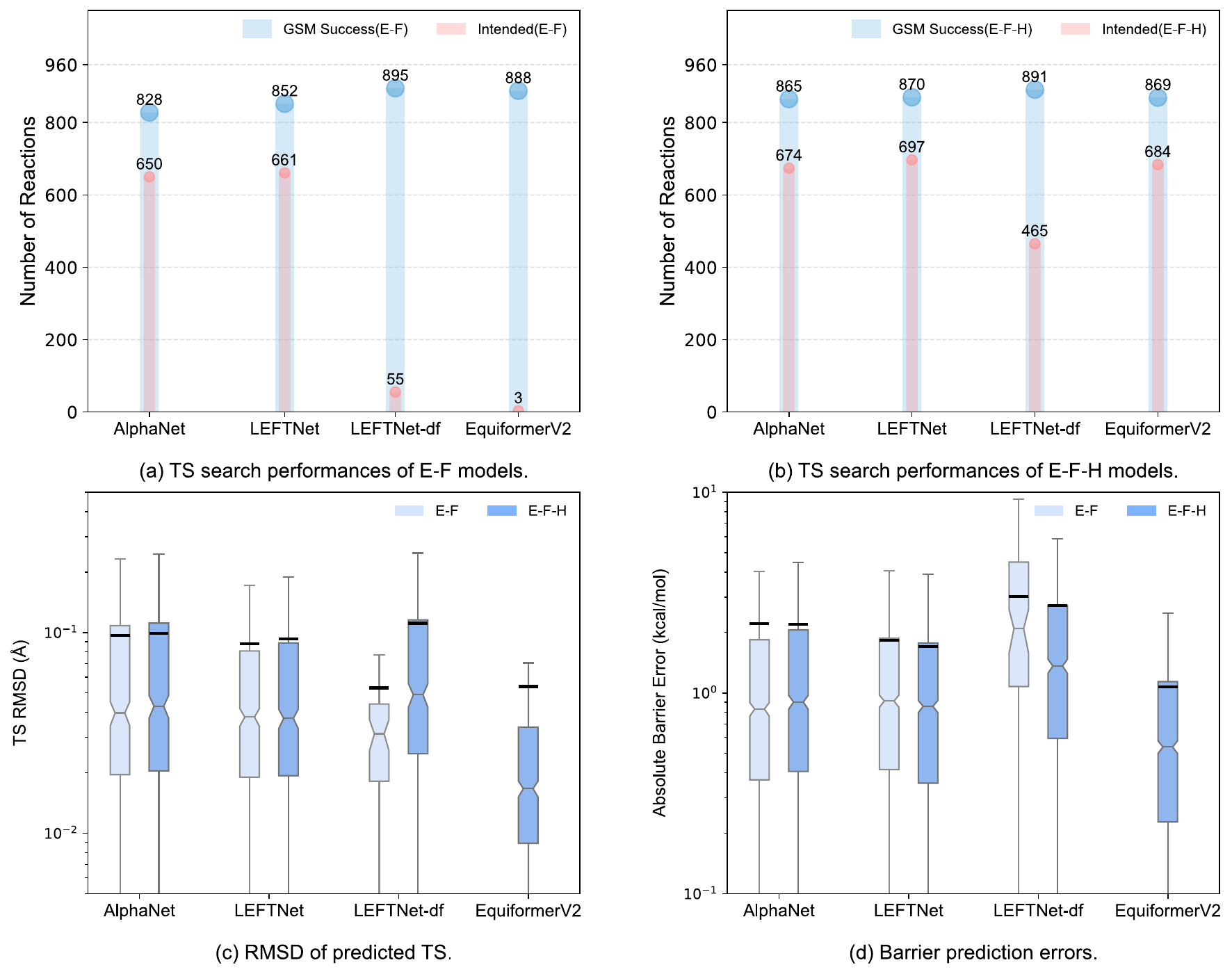}
    \caption{Comparison of MLIP performance in transition state search. (a-b) Number of reactions with a converged GSM pathway (blue) and intended TS (red) with different MLIP models. (c) Root mean square deviation and (d) absolute barrier error to evaluate the quality of intended TSs.}
    \label{fig:ts_search}
\end{figure}

\subsection{Transition State Search Performance}
To evaluate the practical capabilities of reactive MLIPs in realistic TS search scenarios, we assess their performance using our recently developed end-to-end TS search workflow.\cite{zhao2025harnessing} Four key metrics are used for benchmarking: (1) the number of successful GSM calculations, (2) the number of intended TSs—defined by whether the TS connects to the correct reactants and products after the IRC verification, (3) the root-mean-square displacement (RMSD) of optimized TS structures, and (4) the mean absolute error (MAE) of predicted barrier heights, calculated as the energy difference between the MLIP-optimized TS and the reactants (Fig. \ref{fig:ts_search}). GSM success reflects the model’s ability to capture the broader region around the minimal energy pathway (MEP), while the intended TS metric assesses both the quality of the initial guess and the correctness of the local curvature. TS RMSD and barrier MAE quantify the structural and energetic accuracy of the predicted TSs, both of which are essential for reliable kinetic modeling.

These four metrics collectively demonstrate that incorporating Hessian information during training (E–F–H) significantly enhances TS search performance. Among all evaluated metrics, the number of intended TSs showed the most substantial improvement, with EquiformerV2 increasing from just 3 intended TSs under E–F to 684 under E–F–H (Fig. \ref{fig:ts_search}a–b). While the gains observed in autograd-based models were more modest, they remain meaningful given the already strong performance of corresponding E–F models, as demonstrated by LEFTNet, which showed a $\sim 5$\% increase in intended TSs and identified the highest number (697) of intended TSs. Barrier prediction accuracy improved consistently across models, with reductions in barrier MAE of up to 10\% (Fig. \ref{fig:ts_search}d). In contrast, TS RMSD and GSM success rates exhibited minimal improvement, as the Hessian information primarily refines the local curvature of the potential energy surface near the TS, while the TS geometry and GSM calculations are relatively insensitive to marginal improvements in energy and force predictions. Among all evaluated models, EquiformerV2 (E–F–H) achieved the best results in TS RMSD and barrier prediction, with a median TS RMSD of 0.017 \AA~and a median barrier MAE of 0.538 kcal/mol (Fig. \ref{fig:ts_search}c-d). Overall, these results underscore the critical role of Hessian information in improving both the robustness and accuracy of ML-based TS prediction.

\subsection{Performance Overview and Insights}
\begin{figure}[t]
    \centering
    \includegraphics[width=1.0\textwidth]{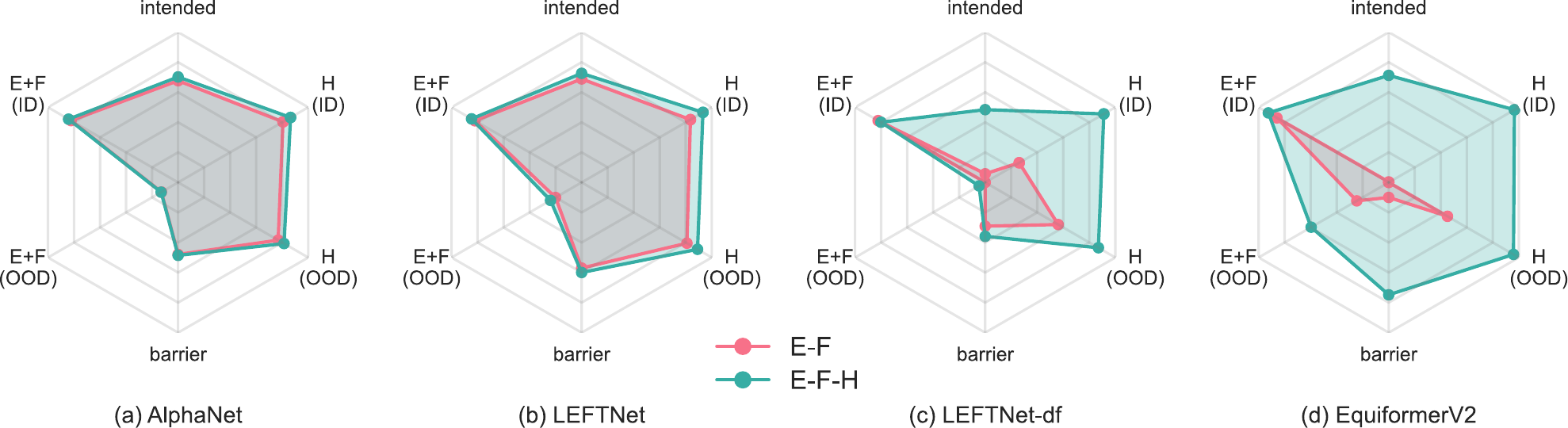}
    \caption{Radar plots comparing the performance of E-F and E-F-H models across different MLIPs.  The six axes represent in-distribution(ID) and out-of-distribution(OOD) MAE of Hessian(H) and energies plus forces(E+F), number of intended TS and barrier prediction error. All metrics are normalized to a 0-100 scale where higher values indicate better performance.}
    \label{fig:performace}
\end{figure}
The overall performances of the four models under E-F and E-F-H loss functions are summarized in four radar plots (Fig. \ref{fig:performace}), which conclude two key general observations:

\textbf{Consistent performance improvement.} The larger hexagonal regions of the E–F–H models (green) compared to the E–F models (pink) highlight the substantial benefits of incorporating Hessian supervision, indicating enhanced predictive accuracy and generalizability. Even for autograd-based architectures (Fig. \ref{fig:performace}a–b), which already demonstrate strong baseline performance, E–F–H training yields notable improvements in both in-distribution and out-of-distribution Hessian predictions, as well as TS search performance. 

\textbf{Breakthrough improvements for direct-force architectures.} While direct-force models have traditionally excelled in energy and force predictions, their application in TS search and long-time scale molecular dynamics simulations has been limited.\cite{zhao2025harnessing,Fu2025LearningSA} Here, we demonstrate that E-F-H training can substantially enhance the capabilities of direct force architectures. This is particularly evident in EquiformerV2 (Fig. \ref{fig:performace}d), which matches LEFTNet—the previously reported best-performing MLIP in TS search across seven MLIP architectures.\cite{zhao2025harnessing} In addition, the E-F-H EquiformerV2 excels in barrier prediction and exhibits remarkable generalization ability, as evidenced by substantial improvements in the E+F(OOD) and H(OOD) metrics.
This can be understood by the fact that MLIPs learn a more exact curvature of PES as more higher-order information is included.

\section{Conclusion}
In this work, we introduced HORM, the largest quantum chemistry database of reactive systems to date, comprising 1.84 million Hessians computed at the $\omega$B97x/6-31G(d) level of theory. By addressing a critical data gap, HORM enables the robust training of reactive machine learning interatomic potentials (MLIPs) with significantly improved Hessian quality—applicable to both direct-force and autograd-based architectures. Notably, when Hessian constraints are used to enforce force symmetry, a representative direct-force model, EquiformerV2, exhibits a 30 to 200 times improvement in Hessian accuracy and TS search performance. This finding highlights a promising direction to overcome key limitations of direct-force MLIPs.

Despite the valuable insights provided by this study, several limitations should be acknowledged. First, HORM primarily consists of small molecular systems—most containing fewer than eight non-hydrogen atoms—which may limit the generalizability of our findings to larger or more structurally complex chemical spaces. Second, the dataset includes a narrow range of elemental species (C, H, O, N) and lacks broad coverage of organic molecular diversity, potentially restricting its applicability across more diverse chemical domains. Ongoing efforts aim to expand HORM to include larger systems and a wider chemical space through the incorporation of elements such as phosphorus, sulfur, and halogens. Third, we have not yet evaluated the long-timescale molecular dynamics (MD) stability of models trained with energy, force, and Hessian (E-F-H) supervision—an important consideration for real-world applications in computational chemistry. Lastly, due to time constraints, we did not fully explore the potential of HORM in training the most effective reactive MLIP; future work should investigate more MLIP architectures and perform comprehensive hyperparameter tuning, such as optimizing the number of Hessian rows used in the loss function and adjusting the E-F-H loss weights.

\section{Acknowledgments}
We would like to thank our entire team from Deep Principle for helpful discussions and support.

\bibliographystyle{unsrt}
\bibliography{Reference}


\appendix
\newpage
\section{Supplementary Materials and Technical Appendices}
\renewcommand{\theequation}{A.\arabic{equation}} 
\renewcommand{\thefigure}{A.\arabic{figure}} 
\renewcommand{\thetable}{A.\arabic{table}}
\setcounter{figure}{0}
\setcounter{table}{0}
\setcounter{equation}{0}
\section*{Supplementary Materials}

\textbf{Hessian Asymmetry Error.} 
In principle, a Hessian matrix should be symmetric. However, in direct-force models, the predicted forces are non-conservative because they are not derived as gradients of a scalar energy function. As a result, the Hessians obtained by differentiating these forces are inherently asymmetric. This asymmetry can directly impact saddle point optimization. To quantify it, we define the asymmetry error as the mean absolute difference between the Hessian and its transpose:
\begin{equation}
\frac{1}{N^2} \sum_{i,j=1}^N\left|H_{i j}-H_{j i}\right|,
\label{eq:asymmertry_value}
\end{equation}

\begin{table}[htbp]
\caption[Table A.1]{Comparison of Hessian Asymmetry Error (eV/\AA$^2$) between HORM-Transition1x validation set and HORM-RGD1 subset.}
\label{tab:asymmetry_comparison}
\renewcommand{\arraystretch}{1.6} 
\centering
\begin{tabular}{lcc}
\toprule
\textbf{Model} & \textbf{Transition1x-val} & \textbf{RGD1} \\
\hline
LEFTNet-df(E-F) & 1.447 (1.271) & 1.184 (1.033) \\
LEFTNet-df(E-F-H) & 0.211 (0.163) & 0.290 (0.253) \\
EquiformerV2(E-F) & 1.235 (1.153) & 1.071 (0.861) \\
EquiformerV2(E-F-H) & 0.073 (0.055) & 0.092 (0.075) \\
\bottomrule
\end{tabular}
\end{table}

\textbf{Computational details.}
All calculations in this study were conducted on the Volcengine computing platform. Transition state (TS) search experiments were executed as bundled single-core jobs on a CPU node equipped with 32 effective cores (Intel Cascade Lake, 2.40 GHz) and 128 GB of memory. Density functional theory (DFT) calculations employed GPU4PYSCF\cite{wu2024gpu4pyscf} as the quantum chemistry engine and were run on NVIDIA A30 GPU cards. MLIP training was also performed on A30 GPUs and H20 GPUs.

\textbf{Hyperparameters.} 
In this work, we evaluate three distinct models: AlphaNet, LeftNet (with both autograd and direct-force variants), and EquiformerV2. The training batch size is selected from \{8, 16, 32\}, depending on the available GPU memory to ensure efficient utilization. AlphaNet employs a 4-layer architecture with a hidden dimension of 128 and 16 attention heads, designed to improve its representational capacity. Both the autograd and direct-force variants of LeftNet are configured with a deeper 9-layer architecture and a larger hidden dimension of 256 to support more complex learning dynamics. EquiformerV2 uses a 4-layer architecture with a hidden dimension of 128, 4 attention heads, and a maximum spherical harmonic degree of $l_{max}=4$, and 4 attention heads. As mentioned earlier, to reduce computational overhead, we randomly sample a subset of columns from each Hessian matrix during training. The number of sampled columns—referred to as the number of reference Hessian rows ($\text{N}_{\text{HR}}$)—is set to 1 for autograd-based models and 2 for direct-force-based models. These hyper-parameters are summarized in the following table.

\begin{table*}[ht]
\centering
\caption{Hyperparameters used for training MLIP models.}
\label{tab:model_config}
\begin{tabular}{lccccccc}
\toprule
\textbf{Model} & \textbf{Layers} & \textbf{Hidden Dim} & \textbf{Heads} &  \textbf{$\text{N}_{\text{HR}}$} & \textbf{Learning Rate} & \textbf{Batch Size} \\
\midrule
AlphaNet       & 4  & 128 & 16  & 1 & $1\times10^{-4}$ & 32   \\
LeftNet (ag)   & 9  & 256 & NA  & 1 & $5\times10^{-5}$ & 64   \\
LeftNet (df)   & 9  & 256 & NA  & 2 & $5\times10^{-5}$ & 64   \\
EquiformerV2   & 4  & 128 & 4   & 2 & $3\times10^{-4}$ & 128  \\
\bottomrule
\end{tabular}
\end{table*}

\end{document}